# Real Space Imaging of the Microscopic Origins of the Ultrahigh Dielectric Constant in Polycrystalline CaCu$_3$Ti$_4$O$_{12}$

S.V. Kalinin, J. Shin, G.M. Veith, and A.P. Baddorf

*Condensed Matter Sciences Division, Oak Ridge National Laboratory, Oak Ridge, TN 37831*

M.V. Lobanov, H. Runge, and M. Greenblatt

*Department of Chemistry and Chemical Biology, Rutgers University, Piscataway, NJ 08854*

**Abstract**

The origins of an ultrahigh dielectric constant in polycrystalline CaCu$_3$Ti$_4$O$_{12}$ (CCTO) was studied using the combination of impedance spectroscopy, electron microscopy, and scanning probe microscopy (SPM). Impedance spectra indicate that the transport properties in the 0.1 Hz – 1 MHz frequency range are dominated by a single parallel resistive-capacitive (RC) element with a characteristic relaxation frequency of 16 Hz. Dc potential distributions measurements by SPM illustrate that significant potential drops occur at the grain boundaries, which thus can be unambiguously identified as the dominant RC element. High frequency ac amplitude and phase distributions illustrate very weak contrast at the interfaces, which is indicative of strong capacitive coupling. These results demonstrate that the ultrahigh dielectric constant reported for polycrystalline CCTO materials are related to the grain boundary behavior.

Recently, $CaCu_3Ti_4O_{12}$ (CCTO) has attracted broad attention due to the discovery of ultrahigh dielectric constant behavior at low frequencies.[1,2,3] A number of attempts to interpret this behavior in terms of intrinsic material properties[4,5] as well as impurity and interface mediated phenomena[6] have been reported. Particularly, recent work by Vanderbilt's group[6] summarized several plausible scenarios for grain boundary mediated low-frequency transport in CCTO. One of the primary explanations for ultrahigh ($\sim 10^5$) dielectric constants in this material is the formation of nonconductive grain boundaries between conductive grains, i.e., Grain Boundary Layer Capacitor behavior, common in doped perovskite titanates.[7] This model for CCTO behavior was recently supported by impedance spectroscopy.[8] However, the universally accepted interpretation of transport behavior in CCTO was not achieved. This can be attributed primarily to the nature of bulk transport measurement techniques, which address only the average transport properties of materials or, in case of impedance spectroscopy, allow the equivalent circuit to be modeled. However, in the general case, the elements of the equivalent circuit cannot be unambiguously associated with individual microstructural elements. Here, we use recently developed scanning probe microscopy (SPM) techniques to directly image the static and frequency-dependent transport behavior of CCTO and demonstrate the origins of ultrahigh dielectric constants.

The $CaCu_3Ti_4O_{12}$ sample was prepared by a conventional solid state reaction with final annealing at 1000 °C in air for a total of 80 hours with two intermediate regrindings. Room temperature synchrotron X-ray diffraction (SXD) data were collected on the SUNY X3B1 beamline at the National Synchrotron Light Source (NSLS), Brookhaven National Laboratory with X-rays of the wavelength 0.6996 Å. The sample was placed in a 1.5 mm glass capillary, which was continuously rotated perpendicular to the scattering plane. The Fullprof program[9] was used for the Rietveld refinement of the crystal structure. The SXD pattern exhibited nearly instrumental resolution with full width at half maximum of 0.010° at 21.8° 2θ and was indexed in an I-centered cubic cell with $a$ = 7.39612(3) Å (Fig. 1 a). No evidence of the presence of secondary phases or crystallographic symmetry lowering was found; neither reflection splitting nor superlattice reflections were observed. The refined atomic coordinates, atomic displacement parameters (ADPs) and relevant interatomic distances are summarized in Table 1. The relatively high values of the profile R-factors reflect primarily the high statistics of the data; consistently, the Le Bail fit with identical values of profile coefficients yielded $\chi^2$=2.52, $R_p$ =11.8%, and $R_{wp}$ =16.2%. Refinement was limited to isotropic ADPs: anisotropic refinement did not yield any significant improvement in the fit quality and components of ADP tensors refined with large uncertainties.

Frequency-dependent macroscopic transport properties of the material were studied with impedance spectroscopy.[10] The sample was polished, and electrodes were fabricated using Pt paste with subsequent annealing. The impedance spectra were acquired at room temperature with a Solartron 1255B impedance analyzer with SI1287 Interface in the frequency range from 0.1 Hz to 1 MHz with excitation amplitudes of 1 Vpp and 50 mVpp. The frequency dependence of the impedance magnitude and phase are shown in Fig. 2. The equivalent circuit is dominated by a single RC element with a resistance of 30.7 MΩ ($R_1$) and capacitance 0.4 nF ($C_1$), corresponding to a dielectric constant κ = 1830. The *RC* relaxation frequency is $\omega_1 = 1/R_1C_1 = 16$ Hz. At higher frequencies, the impedance data indicate the presence of another electroactive element, which can either be attributed to bulk resistance and capacitance or cable inductance. While the frequency range of the spectroscopy is insufficient to achieve an unambiguous conclusion, the corresponding resistance can be estimated as $R_2 \ll 1$ kOhm and the relaxation frequency as $\omega_2 \gg 1$ MHz.

To determine the origins of the frequency-dependent transport behavior in polycrystalline $CaCu_3Ti_4O_{12}$ and correlate the equivalent circuit with the individual microstructural elements, the dc and frequency-dependent transport properties were studied with scanning probe microscopy (SPM). The macroscopic Au electrodes were sputter deposited on the polished surface of the sample using a shadow mask so that the spacing between electrode edges was ~2 mm. Spatially resolved atomic force microscopy and SPM-based dc transport measurements were performed on a commercial SPM system (Veeco MultiMode NS-IIIA) equipped with a custom-built sample holder which allowed *in-situ* biasing. The biases on the electrodes were controlled by function generators (DS 345, Stanford Research Instruments). DC transport measurements were performed in the scanning surface potential microscopy (SSPM)[11] mode with Pt coated tips (NCSC-12 F, Micromasch, $l \approx 250$ μm, resonant frequency ~ 41 kHz) with typical lift heights of 200 nm. The dc current was imposed laterally across the surface by macroscopic electrodes, similar to conventional 4-probe resistance measurements, and the SPM tip is used as a moving voltage-sensing electrode, providing a spatially resolved dc potential distribution image along the surface. Frequency dependent transport measurements were performed in the scanning impedance microscopy (SIM) mode[12,13,14] at 99 kHz with a modulation amplitude of 10 Vpp. Measured in the SIM are the voltage phase and amplitude distribution induced by the ac current applied across the surface, thus providing spatially resolved impedance data. After acquisition of the SSPM images under different bias conditions, the microscope was reconfigured to the SIM mode, thus correlating the dc and ac transport properties of the same region.

The surface topography of a CCTO sample is shown in Fig. 3 a. Due to relatively poor adhesion between individual grains, polishing is associated with the formation of a large number of loose grains and relatively rough surfaces. In this particular case, less than ~20% of the sample surface is suitable for spatially resolved transport measurements, and imaging is limited to small (10-20 μm) scan sizes. Despite this limitation, SSPM and SIM imaging were successfully performed in several locations on the sample surface. The surface potential distribution on a grounded surface is shown in Fig. 3 b. Notice that the presence of a large pore affects surface potential by a relatively small value of ~20-30 mV due to the topographic cross-talk. A number of regions with low (~50 mV) potentials were detected, as indicated by arrows in Figs. 3 b, d. To establish the origins of this contrast, the local chemical composition was determined with an EDAX energy dispersive X-ray spectroscopy device attached to a JEOL JSM-840 scanning electron microscope (SEM). The chemical composition was consistent at all points, and no evidence for a secondary phase formation was observed (Fig. 1 b). The dark spots observed on the SEM images disappeared after prolonged (~30 sec) e-beam exposure and therefore may be attributed to surface contaminants that can locally affect the local work function of the material as seen on the SSPM image (Fig. 3 a).

The surface potential of the same region under lateral 10 V bias is shown in Fig. 3 c. Note that the surface potential exhibits well-defined potential drops within the image. The direction of the drops inverts with the bias, as illustrated in Fig. 3 d. The magnitude of the potential drops is ~30-50 mV, comparable with that expected for a 10 V bias applied across ~2 mm for an average grain size of ~10 μm. This behavior indicates a significant contribution of grain boundary resistance to the dc transport in the material. Combined with the impedance data, these results show unambiguously that the low frequency RC element in the equivalent circuit is due to grain boundary behavior.

Corresponding SIM amplitude and phase images at 99 kHz are shown in Figs. 3 e and f, respectively. The amplitude image shows significant topographic cross-talk, which is expected since the SIM amplitude signal is directly proportional to the tip-surface capacitance gradient.[12] The SIM phase data shows only minor phase variations at some interfaces. This is expected, since at high (~99 kHz) frequencies the impedance is more than 3 orders of magnitude lower due to capacitive coupling across the interfaces. Hence, the ac amplitude drops at the electrodes and becomes significant in the bulk and the SIM amplitude images do not exhibit amplitude contrast at the interfaces. At low frequencies, the amplitude images correspond to the dc transport measurements by SSPM, and the resistive component of the grain boundary impedance dominates the image. The phase signal, $\varphi$, in the high-frequency regime scales as $\tan(\varphi) \sim \omega^{-1}$, and is also small for $\omega >> \omega_1 = 16$ Hz.

To summarize, dc and frequency-dependent transport in polycrystalline $CaCu_3Ti_4O_{12}$ with a giant dielectric constant was studied using the combination of impedance spectroscopy, electron microscopy, and SPM. The synchrotron X-ray and SEM data indicate that the sample is free of impurities and the crystallographic structure agrees with the one previously reported from a neutron diffraction study.[15] Impedance spectra indicate that the transport properties in the 0.1 Hz – 1 MHz frequency range are controlled by a single parallel RC element with a characteristic relaxation frequency of 16 Hz. Spatially resolved dc transport measurements by SSPM illustrate that significant dc potential drops occur at the grain boundaries, which thus can be unambiguously identified as the dominant RC element. Moreover, impedance images acquired at high frequencies illustrate very weak grain boundary contrast, which illustrates capacitive coupling across the interfaces. These spatially resolved results demonstrate that the ultrahigh dielectric constant reported for polycrystalline CCTO materials are related to grain boundary behavior.

Research was performed as a Eugene P. Wigner Fellow and staff member at the Oak Ridge National Laboratory (ORNL), managed by UT-Battelle, LLC, for the U.S. Department of Energy under Contract DE-AC05-00OR22725 (SVK). Support from ORNL Laboratory Research and Development funding is acknowledged (SVK and APB). GMV was supported in part by an appointment to the ORNL Postdoctoral Research Associates Program administered jointly by the Oak Ridge Institute for Science and Education and ORNL. The work at Rutgers was supported by the NSF Grants DMR 99-07963 and DMR 02-33697. The research carried out at the NSLS at Brookhaven National Laboratory was supported by the U.S. Department of Energy, Division of Materials Sciences and Division of Chemical Sciences. The SUNY X3 beamline at NSLS was supported by the Division of Basic Energy Sciences of the U.S. Department of Energy under Grant No. DE-FG02-86ER45231. The authors are grateful to C. Botez and P. W. Stephens for the help with SXD data collection and G. Ownby, J. Luck, and Dr. P. Khalifah (ORNL) for invaluable help in sample preparation.

**Table 1.**

Fractional coordinates, isotropic ADPs ($B_{iso}$) and selected interatomic distances (Å) refined by the Rietveld method for the $CaCu_3Ti_4O_{12}$ sample. Space group $Im\overline{3}$, $a$ = 7.39612(3)Å, Z = 2, $\chi^2$ = 2.64, $R_p$ = 12.0%, $R_{wp}$ = 16.5%, $R_B$ = 4.35%, $R_F$ = 3.84%.

| Atom | Site | x | y | z | $B_{iso}$ (Å²) |
|---|---|---|---|---|---|
| $Ca^{+2}$ | 2a | 0 | 0 | 0 | 0.78(9) |
| $Cu^{+2}$ | 6b | 0 | ½ | ½ | 0.38(6) |
| $Ti^{+4}$ | 8c | ¼ | ¼ | ¼ | 0.31(7) |
| $O^{-2}$ | 24g | 0 | 0.1806(9) | 0.3031(8) | 0.38(9) |
| Distances: | | | | | |
| Ca-O (×12): 2.609(3) | | | | | |
| Cu-O (×2): 1.977(4), 2.775(6), 3.256(5) | | | | | |
| Ti-O (×6): 1.959 (3) | | | | | |

**Figure Captions**

**Figure 1.** (a) Experimental, calculated and difference patterns for the Rietveld refinement of room temperature synchrotron X-ray diffraction data. Inset shows a schematic representation of the crystal structure of $CaCu_3Ti_4O_{12}$. (b) SEM image of polished CCTO surface. Inset shows well-polished regions similar to that used in SPM transport measurements.

**Fig. 2.** Impedance magnitude (a) and phase angle (b) as a function of frequency. The inset shows a typical equivalent circuit with grain boundary and bulk contributions. Note that without additional information the elements of equivalent circuit cannot be unambiguously associated with individual microstructural elements such as grain boundaries, bulk, or contacts.

**Fig. 3.** (a) Topography and (b) potential of the grounded CCTO surface. Surface potential of the sample laterally biased by + 10 V (c) and -10 V (d). Scanning impedance amplitude (e) and phase (f) images of the same region. Scale is 500 nm (a), 100 mV (b, c, d), and 2° (f).

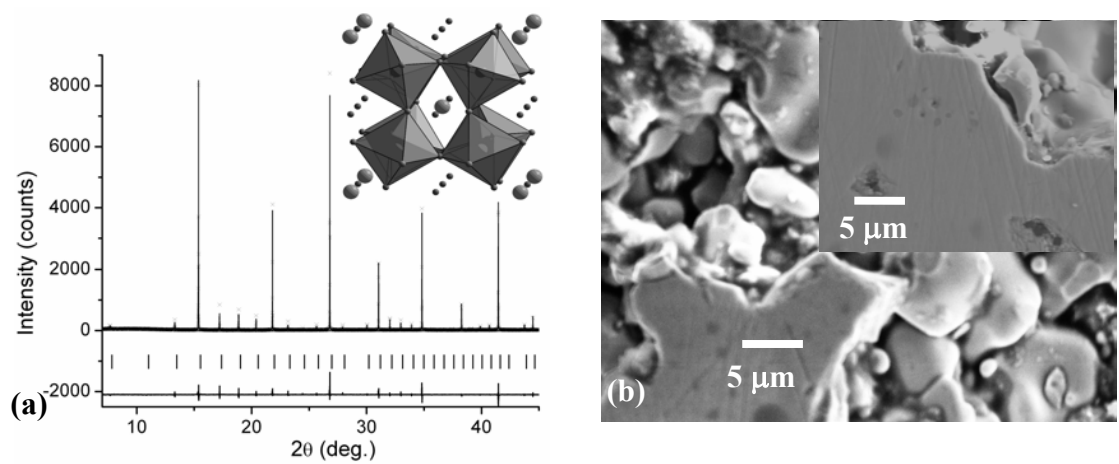

**Fig. 1.** S.V. Kalinin, J. Shin, A.P. Baddorf et al.

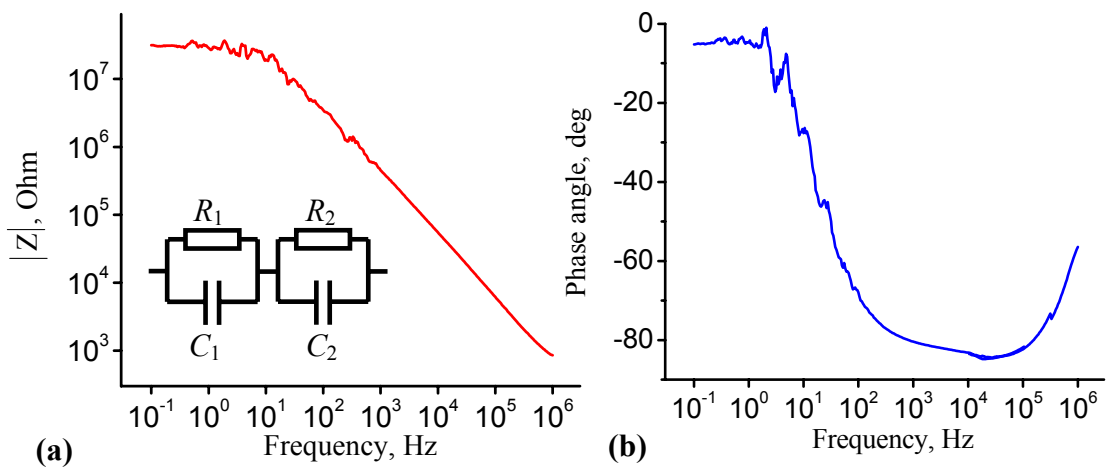

**Fig. 2.** S.V. Kalinin, J. Shin, A.P. Baddorf et al.

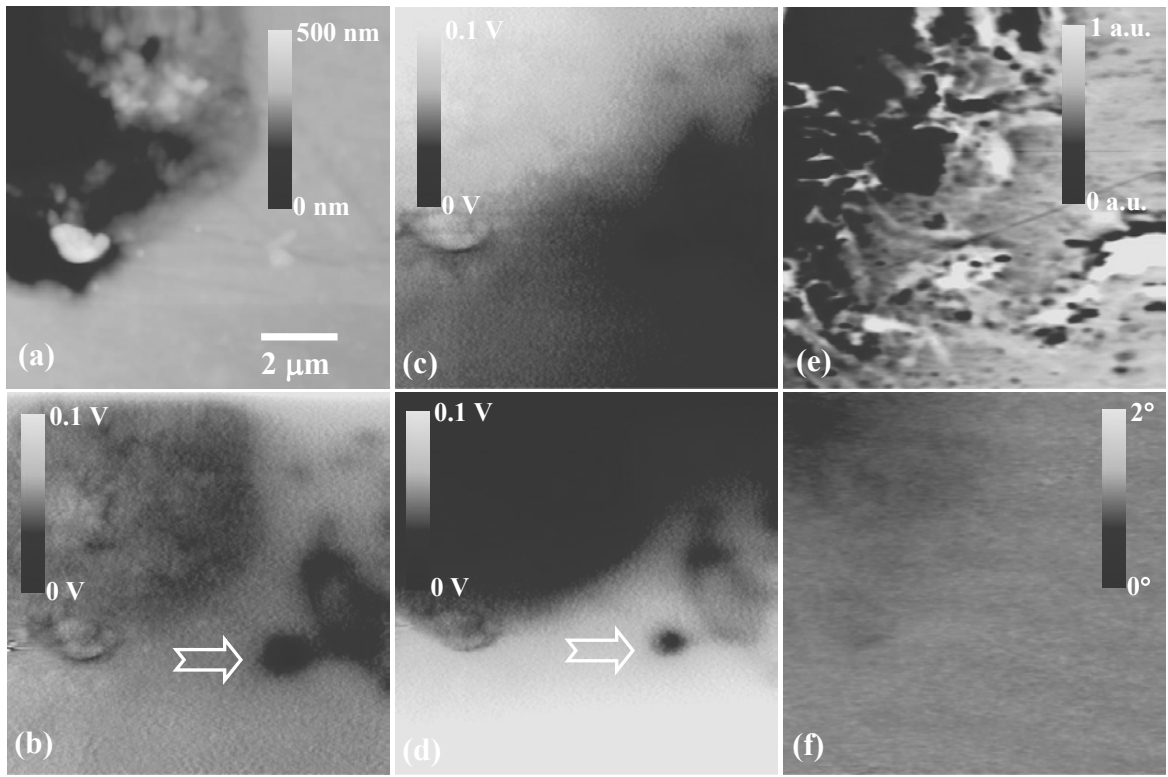

**Fig. 3.** S.V. Kalinin, J. Shin, A.P. Baddorf et al.